\documentclass[amsmath, amssymb, aps, pra,reprint, twocolumn,longbibliography,floatfix,nofootinbib]{revtex4}
\usepackage[utf8]{inputenc}
\usepackage{amsmath}
\usepackage{amssymb}
\usepackage{xcolor}
\usepackage{comment}
\usepackage{titlesec}
\usepackage[colorlinks]{hyperref}
\usepackage{amsmath,amssymb}
\usepackage{bbold}
\usepackage{float}
\usepackage{graphicx}
\usepackage{dcolumn}
\usepackage{bm}
\usepackage{dsfont}
\usepackage{url}
\usepackage{mathtools}

\begin{document}
\title{Solving workflow scheduling problems with QUBO modeling}

\author{A. I. Pakhomchik}
\address{Terra Quantum AG, St.\,Gallerstrasse 16A, 9400 Rorschach, Switzerland}

\author{S. Yudin}
\address{Terra Quantum AG, St.\,Gallerstrasse 16A, 9400 Rorschach, Switzerland}

\author{M. R. Perelshtein}
\address{Terra Quantum AG, St.\,Gallerstrasse 16A, 9400 Rorschach, Switzerland}

\author{A. Alekseyenko}
\address{Volkswagen Group of America San Francisco, CA, USA}

\author{S. Yarkoni}
\address{Volkswagen Data:Lab, Munich, Germany}

\begin{abstract}
In this paper we investigate the workflow scheduling problem, a known NP-hard class of scheduling problems. We derive problem instances from an industrial use case and compare against several quantum, classical, and hybrid quantum-classical algorithms. We develop a novel QUBO to represent our scheduling problem and show how the QUBO complexity depends on the input problem. We derive and present a decomposition method for this specific application to mitigate this complexity and demonstrate the effectiveness of the approach. 
\end{abstract}
     
\maketitle

\section{Introduction}
\label{sec:intro}
Quantum computing has garnered increased attention in recent years, in both industrial and academic contexts. In general, the aim is to develop specialized hardware that can be programmed to simulate a quantum mechanical process, which is classically intractable~\cite{benioff_computer_1980,feynman_simulating_1982}. Construction of algorithms using quantum bits (qubits) currently proceeds as multiple paradigms, the most well-known of which being the gate model~\cite{Feynman1986} and the adiabatic quantum computing~\cite{farhi_AQC_2000} model. In the former, unitary operators are used to manipulate individual qubits' states to construct the logical operators. In the latter, a system is initialized in a simple superposition of all possible states, and slowly evolved to represent a final function (also called a final Hamiltonian). It has been shown that these models are polynomially equivalent~\cite{aharonov2008}, and both have been studied extensively.

Advancements in the development and production of quantum hardware has led to the manufacture of quantum hardware prototypes of various sorts, often made publicly-accessible. Companies such as Google~\cite{Arute2019}, IBM~\cite{ibm65qubits}, and D-Wave Systems~\cite{johnson_quantum_2011}, among others, all offer cloud-based access to a suite of quantum algorithms tailored for their respective quantum processing units (QPUs). The purpose of these is to exploit such quantum algorithms in order to address computationally difficult, and sometimes intractable, problems in fields such as machine learning, molecular and physical simulation, and optimization. Of these, significant work has already been done in the realm of optimization, largely due to the implementation of quantum annealing~\cite{LantingEntanglement2019} and quantum approximate optimization algorithm (QAOA)~\cite{farhi_quantum_2014}. Both of these are metaheuristic quantum optimization algorithms which can be implemented in currently-available quantum hardware. Previous literature highlight the efforts to construct suitable optimization problems that can exploit these quantum algorithms in both academic~\cite{Lucas2014} and industrial~\cite{tfo,denso,mcps,YarkoniQA2021,bmw} circles. 

How exactly quantum algorithms can impact combinatorial optimization in the absence of error-correction remains an open question. Furthermore, there is little evidence of concrete use of quantum algorithms for real-world applications, outside of a select choice of showcase examples (e.g.,~\cite{quantum_shuttle}). While error-correction will allow implementation of provably asymptotically faster quantum algorithms with respect to their classical counterparts (such as Shor's factoring~\cite{Shor1997}, Grover's search algorithms~\cite{Grover1996}, and solving systems of linear equations~\cite{HHL,Perelshtein2020}), it is unknown if noisy intermediate-scale quantum (NISQ~\cite{Preskill2018quantumcomputingin}) computing can overcome its limitations to provide similar value. In the meantime, hybrid quantum-classical algorithms have emerged to bridge the gap until the end of the NISQ era is reached. Construction of variational algorithms has been demonstrated, in particular for gate-model quantum computers, to perform specific tasks in quantum machine learning~\cite{Skolik2021}, quantum chemistry~\cite{Kandala2017}, and optimization~\cite{Harrigan2021}. In this paper, we compare such hybrid algorithms to a variety of techniques to solve a specific class of scheduling problems, the \textit{workflow scheduling} problem.

The rest of this paper is organized as follows: Section~\ref{sec:applications} introduces the concepts behind the different scheduling problems, as well as the previous works studied in quantum computing. Section~\ref{sec:formulation} formally introduces the version of workflow scheduling investigated in this paper, and develops the methods required to model this problem as a QUBO for quantum optimization algorithms, including a decomposition technique for solving large QUBOs. In Section~\ref{sec:benchmarking} we present the data used to generate the problem instances, and the algorithms used to solve them in experiments. The results are presented and discussed in Section~\ref{sec:results}, and our conclusions are presented in Section~\ref{sec:conclusions}.

\section{Applications of scheduling problems}
\label{sec:applications}
Many applications related to supply chain and logistics optimization can be formulated as certain classes of scheduling problems. Typically, these problems can be described as a set of jobs (composed of individual sequences of operations), each taking a non-negative amount of time, that must be completed in the minimum amount of time (known as the makespan) on a set of machines. There are different variants of the job-shop scheduling problems~\cite{Lawler:78:Sequencing-jobs, LenstraRinnooy-Kan:78:Complexity-of-scheduling, DuLeung:91:Scheduling-chain-structured, BevernBredereck:16:Precedence-constrained-scheduling, LenstraRinnooy-Kan:77:Complexity-of-machine, GareyJohnson:78:Strong-NP-completeness}, each with their own set of constraints and conditions that uniquely define them. Dynamic resources, supply constraints, time windows (and more), all are examples of constraints that may be used to tailor a sub-class of scheduling for a particular interesting case. A particularly general and well-known version of the problem, the job-shop scheduling problem (JSP), typically refers to the case where there are $N$ jobs to be executed on $M$ machines, and no other additional constraints. This simple version of the problem is already NP-hard; a well-known Ising model formulation has been used to study the JSP in the context of quantum computing~\cite{jsp}. Another example is a similar scheduling problem, the Nurse Scheduling problem, which attempts to schedule nurses to shifts based on personal availability and other hard constraints~\cite{Ikeda2019}. In this work we motivate one specific subclass of scheduling, namely the \emph{dynamic resource workflow scheduling problem}. The problem we consider is motivated by a real-world use-case in the automotive industry, the quality control testing of manufactured cars at the end of an assembly line. After a car is produced, a sequence of tests and checks are performed by workers on the factory floor to ensure the quality of production. This particular problem has the following constraints: for a set of tests to be performed, some tests may have sub-tasks that are conflicting with other tests; the number of workers available changes over time; and most importantly, some tests may be dependent on others to be completed first. The objective of the optimization problem is therefore to determine the sequence of tests that minimizes the total amount of time required to complete all the tests (i.e., minimize the makespan). 

\begin{figure*}
     \noindent\centering{
    \includegraphics[width=140mm]{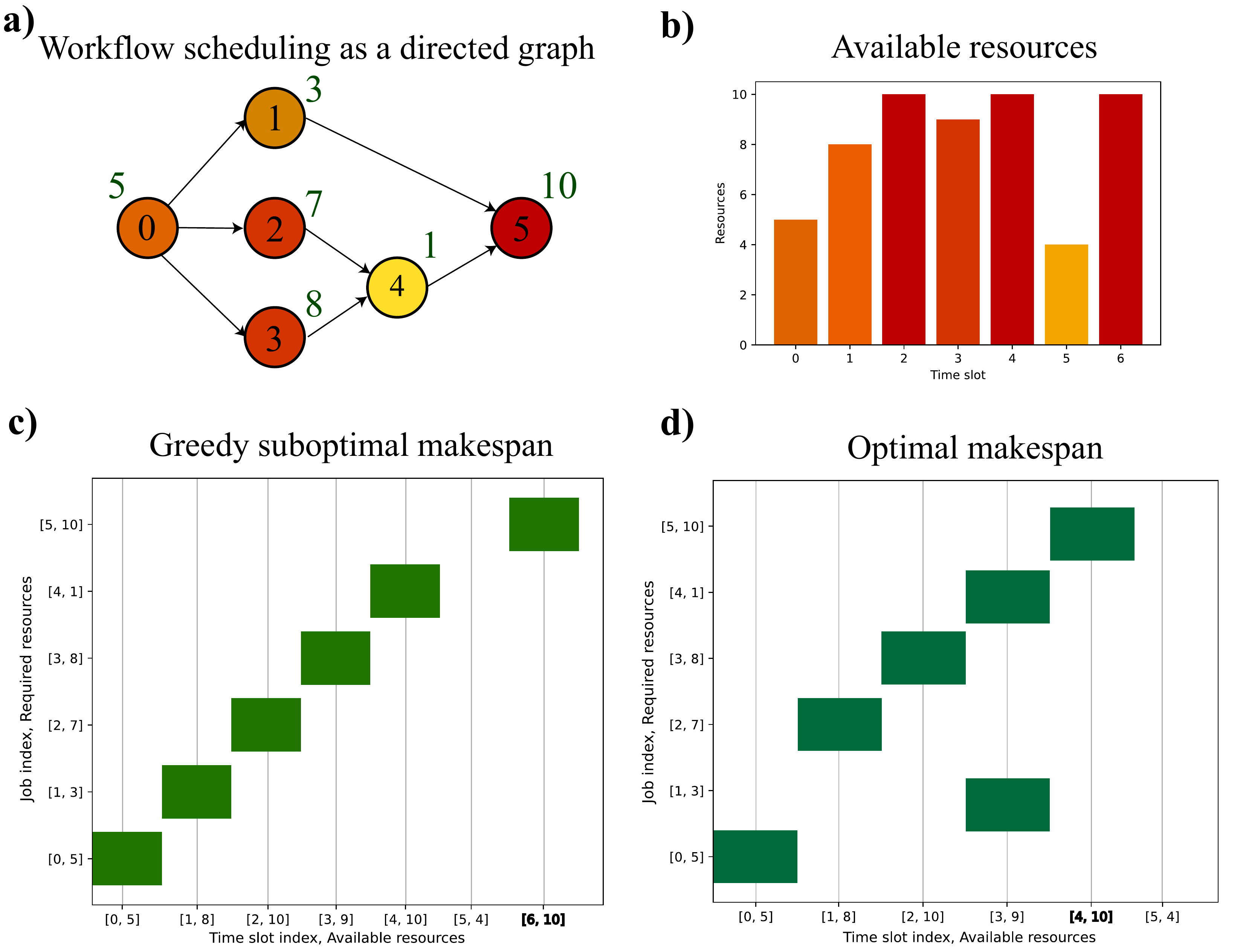}
    }
    \caption{ 
    {\bf Example of a workflow scheduling problem with 6 jobs. }
    a) Directed acyclic graph with 6 nodes (job index is inside the circle) that illustrates the required ordering (parents-children relations) and resources for each job (number next to the circle).
    b) Available resources for each time slot when one or more jobs can be completed. 
    The maximum number of time slots is fixed to be seven.
    c) Makespan map of the problem processed by sub-optimal greedy algorithm.
    Map shows at which time slot which job should be completed.
    The total makespan for greedy solution is 7 time slots.
    d) Makespan map of the problem processed by optimal algorithm.
    The total makespan is 5 time slots.
    Two jobs \#1 and \#4 are completed in a same time slot \#3 since their parents were completed and the number of available resources (9) is higher than the required resources for both jobs (3+1).
    }
    \label{5nodes}
\end{figure*}

We define the problem formally as follows: given a set $J = \{J_1, \cdots, J_N\}$ denoting $N$ jobs (we consider the case where each job has one operation), each job takes an amount of time $T(J_i)$ and requires at least $R(J_i)$ workers to be started and executed. The set of dependencies for each job is $\mathcal{D}(J_i) = \{J_j, J_k, \cdots\}$, denoting all tasks which are dependent upon the completion of $J_i$. Lastly, the vector $\vec{W}$ represents the number of available workers to perform the tasks at each time step. We consider workers (i.e., the available resources to perform tasks at each step) as identically qualified and thus they are able to work on any task in the workflow scheduling problem. In general, this is not necessarily the case, and one could extend the models we derive to accommodate for multiple categories of workers (where tasks also depend on different or even multiple categories) seamlessly. Visually, workflow scheduling can be represented as a directed acyclic graph (DAG), where jobs $J_i$ are represented as nodes and edges illustrate the set of dependencies for each job. An example of a 6-node problem with limited available resources at each time slot is shown in Fig.\,\ref{5nodes}a,b. The solution of a problem is a makespan map that shows when each job is completed. Sub-optimal and optimal makespan maps for this example problem are presented in Fig.\,\ref{5nodes}c,d respectively. 

\section{Problem Formulation}
\label{sec:formulation}

Before formulating the workflow scheduling problem as a QUBO, we introduce some assumptions that allow for the simplification of the optimization, without loss of the generality.
Here, we assume that (i) a single job can not occupy more than a single time slot, while the length of a job equals exactly one time slot length; (ii) multiple jobs can be started in a single time slot if there are sufficient resources; (iii) all resources are identical and the amount of available resources at the current time step is bounded by the number $r_{\text{max}}$; (iv) jobs can only be started if all parent jobs are completed. 

\subsection{Binary Optimisation} 
\label{bin_opt}

In a binary variable formulation, the solution of a workflow scheduling problem is represented by the set of decision variables $\vec{x}$, where each binary variable $x_{i,t}$ represents the following:

\begin{equation}
    x_{i,t}=\begin{cases}
            1, & \text{if $i$th job is  started  in the $t$th time slot,}\\
            0, & \text{otherwise.}
    
            \end{cases}
\end{equation}

Using this notation, we can express all the necessary constraints which appear naturally in the workflow scheduling.

{\it A job is started only once.\quad} All jobs start once and only once, otherwise we introduce unnecessary repetitions that use more resources than needed. This constraint is represented by a simple equality:

\begin{equation}\label{only_once}
    \sum_{t}x_{i,t}=1,~\forall~i.
\end{equation}\\

{\it All jobs are started in order. \quad}
We must ensure that no solutions to the QUBO allow starting a task before the previous dependent tasks are completed. In order to satisfy this condition, we introduce a constraint in the following way. Let us denote the set of all children for $i$th job as $O_{i}$. In our binary formulation we introduce the following penalty:

\begin{equation}
    x_{i,t_{1}}x_{j, t_{2}}=0\quad\forall i, t_{2}\leqslant t_{1}, j\in O_{i}.
    \label{order}
\end{equation}

This is sufficient (along with Eq.\,\eqref{only_once}) to ensure causality of dependent tasks. Any ordering of tasks in which children are scheduled before their parents result in higher objective value than the correct ordering, which is what we require. \\

{\it A job is started if and only if there are enough resources. \quad}
A job starts in the $t$th time slot only if the amount of resources related at $t$th time slot is enough to cover the job. 
Practically, such a model is appropriate under the conditions of identical resources whose availability fluctuates given a known schedule. To encode this constraint, we use the following system of inequalities:
\begin{equation}
\sum_{i}x_{i,t}r_{i}\leqslant r_{t},\quad\quad r_{i}, r_{t}\in\left[0, r_{\text{max}}\right],
\end{equation}
where $r_i$ is the amount of resources required by the $i$th job, $r_t$ denotes the available resources at time slot $t$, and $r_{max}$ is the total amount of resources in the problem.\\

Giving formulation of all the constraints in a binary format, we can construct a single quadratic cost function containing constraints as additive penalties, i.e. the QUBO format.

\subsection{Constructing the QUBO formulation}

\subsubsection{Transforming inequality to equality }

Inequalities are transformed into equalities for binary variables by introducing binary slack variables in the following manner:
\begin{align}
    \sum_{j}a_{i,j}x_{j}\leqslant b_{i}\Longleftrightarrow b_{i}-\sum_{j}a_{i,j}x_{j}
    = \sum_{k=0}^{\mathcal{N}_{i}-1} \alpha_{i,k} 2^{k}.
    \label{slack_variables}
\end{align}

Here, $\mathcal{N}_{i}=\lfloor \left(\log_{2}\left(D_i\right)\right) \rfloor+1$ with $D_i=\max(b_i-\sum_{j}a_{i,j}x_j)$, and $\lfloor...\rfloor$ means rounding for positive integers and 0 otherwise. In the case of negative $D_i$, there is no $x$ that satisfies the inequality.
If $\sum_j a_{i,j} x_j = b_i$, then $D_i = 0$, resulting in $\mathcal{N}_i = 1$.

\subsubsection{Objective function}
It is important to note that our workflow optimization formulation focuses on solving the NP-hard makespan minimization problem, rather than the NP-complete decision problem. Thus, we define the objective cost as a function of the makespan, which is to be minimized. We introduce a penalty term which penalizes starting a job after the expected runtime, whose magnitude is a tuned hyperparameter. The resulting objective has the form:

\begin{equation}
    \widetilde{C}=\sum_{i,t>R}f\left(t-R\right)x_{i,t},
    \label{obj_cost}
\end{equation}

where $f$ is a monotonically increasing function of $t-R$ with $R$ being the total expected runtime. Combining all constraints in the form of penalties, we obtain
\begin{multline}
    C=\widetilde{C}+\beta\sum_{i}\left(\sum_{t}x_{i,t}-1\right)^2+\gamma\sum_{i,t_1,t_2\leqslant t_1,j\in O_{i}}x_{i,t_{1}}x_{j,t_{2}}+\\
    + \epsilon\sum_{t}\left(\sum_{i}x_{i,t}r_{i}+\sum_{k=0}^{\mathcal{N}_t-1}\alpha_{t,k}2^k-r_t\right)^2,
\end{multline}

where $\beta, \gamma, \epsilon$ are penalty weights for one-time job start, ordering, and limited resources, respectively. Here, $\mathcal{N}_t=\lfloor\log_{2}\left(r_{t}-\sum_{i}x_{it}r_{i}\right)_{\text{max}}\rfloor+1$ is the number of slack variables for the \textit{t}th time-step as per Eq.\,\eqref{slack_variables}.

Unbounded search could be performed to find optimal penalty weights that maximize the probability of obtaining minima, but we set $\beta=\gamma=\epsilon=A,~f(t-R)=t-R$, disregarded the rigorous analysis of the behaviour of solvers for different values of hyperparameters. 
Thus, QUBO could be represented as the sum of two terms:

\begin{equation}\label{qubo_two_terms}
    C = \widetilde{C} + AQ_0.
\end{equation}

The guarantee that an optimal solution would be feasible is similar to the estimate in~\cite{srp}, where such sufficient conditions for feasibility were found. Specifically, $A>\widetilde{C}\left[feas\right],$ where $feas$ is any feasible solution to the problem. Indeed, supposing the optimal solution $opt$ is not feasible in this case, we get a contradiction by the following chain of inequalities:

\begin{equation}\label{ineq_for_weight}
    \widetilde{C}\left[opt\right] + AQ_0\left[opt\right]\geqslant A + \widetilde{C}\left[opt\right]\geqslant A > \widetilde{C}\left[feas\right].
\end{equation}

\subsubsection{Size reduction}
Lastly, given that we know the resource distribution beforehand -- both required and available -- we simplify the problem by assuming that the $i$th job can not be started at $t$th time slot if there are not enough resources:

\begin{equation}
    x_{i,t}=0\quad\quad if\quad r_{i}>r_t.
\end{equation}

This trick allows us to reduce the problem size by conditioning on infeasible variables rather than adding penalties for the inability to start a job. 

\subsection{Decomposing the QUBO formulation}
\label{decomposition_section}

Combining all constraints into a single objective function-- including all ancillas necessary for transforming inequalities-- generates a significant increase in the number of variables in the final QUBO. This signifies the polynomial overhead incurred by transforming generic optimization problems to QUBO forms. However, we can simplify the problem using decomposition techniques, transforming a larger QUBO into a set of smaller instances. We accomplish this by leveraging the hierarchical structure of the workflow illustrated by a strict parents-children relation, as dictated by the individual tasks' dependencies. These smaller instances (sub-problems) are created in a way that ensures all constraints in the larger problem remain satisfied.
Such a decomposition significantly simplifies the problem complexity, and is especially useful at larger problem sizes since problems with hundreds of jobs are challenging even in the most efficient linear programming formulations. The complete optimisation of the problem is therefore performed by solving these sub-problems in a dynamic manner. Interestingly, such a method is applicable not only for quantum algorithms but also for any other exact or heuristic discrete optimisation tool or formulation, including LP, QUBO, HOBO, etc. We now describe the method in more detail.

We start with finding the roots of the directed graph representation of the workflow scheduling problem (i.e., jobs without any parents), and a fixed number of their descendants, $m$ jobs in total. We also fix the number of time slots $n$ that can be processed in a single sub-problem. 
Therefore, we have to schedule $m$ jobs across $n$ time slots. In other words, $n, m$ are now hyperparameters that control the globality of each of the sub-problems.

In order to formulate such sub-problems as QUBO correctly we slightly modify the constraints.
Firstly, we relieve the requirements on all jobs to be started only once that are stated in Eq.\,\eqref{only_once}. Instead we set a constraint that allows the completion of a job either zero or one time:
\begin{equation}\label{greedy_once}
    \sum_{t} x_{i,t} \le 1~\mbox{for all}~i
\end{equation}
Such a constraint comes from the local uncertainty of how many jobs have to be completed in the corresponding sub-problem. For this purpose, we rewrite the cost term from Eq.\,\eqref{obj_cost} in the following way:
\begin{equation}\label{greedy_objective}
    \widetilde{C} = - \sum_{i, t}^{m, n} x_{i,t},
\end{equation}
which encourages the completion of more expensive jobs in terms of resources.
Secondly, the order constraint set in Eq.\,\eqref{order} is not suitable anymore since it does not penalize the case where a child with uncompleted parent was started.
Therefore, to address this issue, we change the cost function term for order violation from Eq.\,\eqref{order} in the following way
\begin{equation}
    \, x_{j,t_{1}}\left(1 - \sum_{t_2 < t_1} x_{i,t_{2}}\right) \quad\quad\forall i,\quad\forall t_{1}, \quad\forall j\in O_{i}.
\end{equation}
The minimum of this is now when $x_{j,t_1}=1$ and any $x_{i,t_2}=1$ (where order is conserved), or if $x_{j,t_1}=0$ and so no child task of $i$ is scheduled. Using the solution of this sub-problem, we can define new roots and their descendants, which are considered as the next sub-problem until all jobs are completed in this manner.

\begin{figure}[h!]
     \noindent\centering{
    \includegraphics[width=82mm]{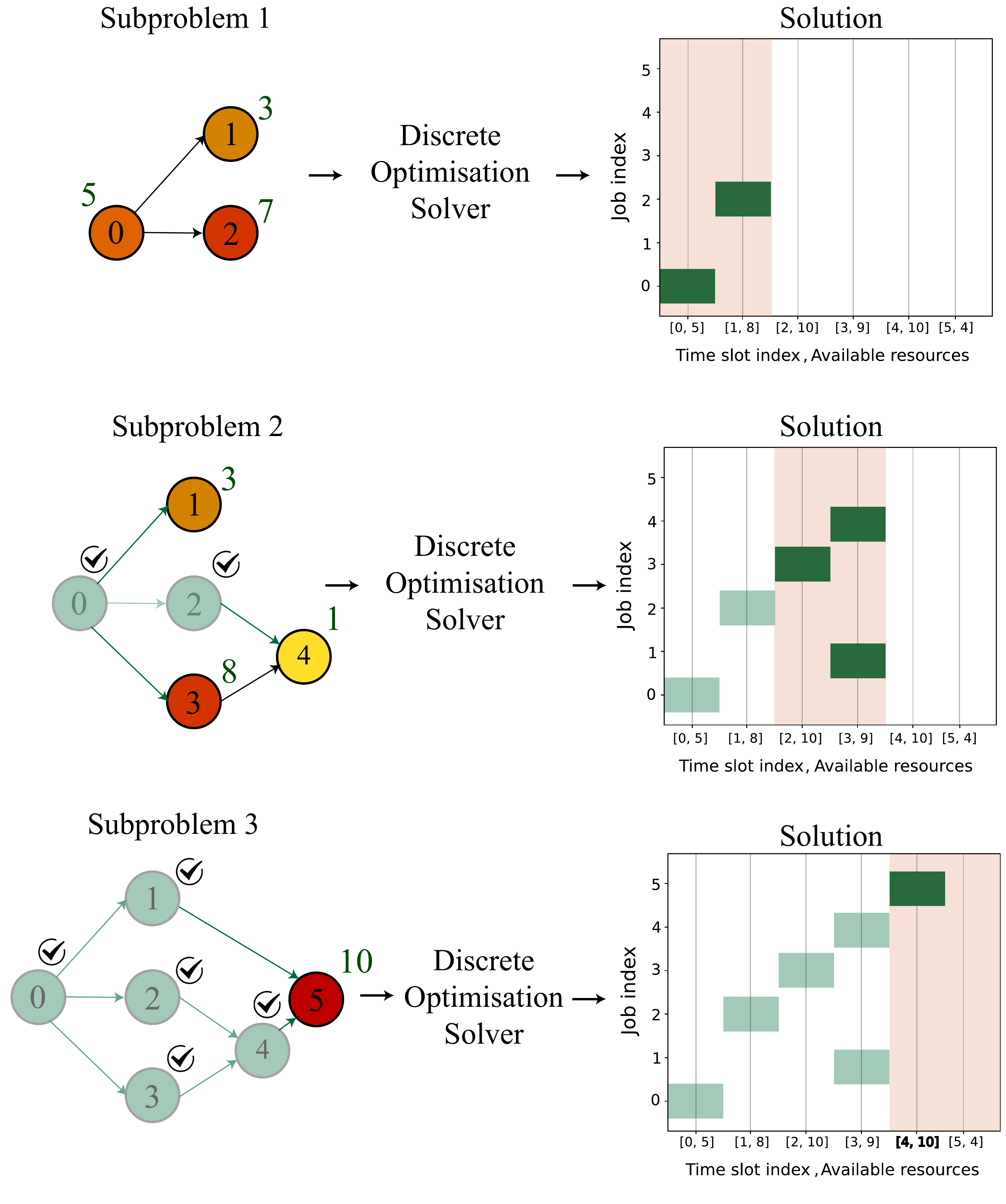}
    }
    \caption{ 
    {\bf Decomposition technique for the problem depicted in Fig.\,\ref{5nodes}.}
    According to the algorithm, the whole problem is divided into three subproblems where we aim to complete at max three jobs in two time slots and create the whole makespan map (depicted on the left). 
    In the first subproblem, jobs $0$ (root), $1$ (first descendant), $2$ (first descendant) are available for processing, however only $0$ and $2$ are completed due to the lack of resources: $0$ is completed since it is a root and $2$ because it requires more resources than $1$.
    In the second subproblem, jobs $1$ (new root), $3$ (new root), $4$ (first descendant) are considered and all three are completed.
    In the last subproblem, only job $5$ (new root) remains and it is completed in a single time slot.
    Each subproblem can be solved in any suitable optimisation formulation via any discrete solver.
    }
    \label{decomposition}
\end{figure}

To illustrate the decomposition method let us consider the 6-node example depicted in Fig.\,\ref{5nodes} from before.
Here, we set the number of jobs to be $n=3$ and number of time slots $m=2$ in a single subproblem.
The scheme of the subproblems division and their solutions is shown in Fig.\,\ref{decomposition}.
Here, the first three jobs are picked since the job with index $0$ is a root and $1$ and $2$ are its closest descendants. 
It is impossible to place all three jobs in two time slots, therefore only job $0$ and $2$ are completed: $0$ \textit{must} be completed since it is a root, and $1$ requires more resources than $2$.
More expensive jobs are completed earlier, if possible, since we do not know if there are enough available resources in the next time slots for these jobs.
Within the second subproblem, jobs $1$ and $3$ are new roots since their ancestor is completed, and job $4$ is the closest descendant for job $3$. In contrast with the first subproblem, all three jobs are completed. On the last subproblem, only job $5$ is left and is successfully completed in a single time slot.

The advantage of such an decomposition algorithm lies in its dynamic nature. 
In case of failures during the problem solution when the number of available resources is changed, the algorithm can process such an unexpected event -- there is no need to restart the whole solution construction. However, the disadvantage of the presented algorithm is its locality, and therefore, the fact that it may provide sub-optimal solutions.
One way to further minimize the total makespan (and ultimately perform global optimisation) is to increase the size of subproblems. 
The worst case scenario requires the subproblem to be the same size as the whole problem.
Existing high-performance linear programming solvers, e.g. CPLEX, struggle to schedule more than 40 jobs in a reasonable amount of time, and thus, limit the maximum problem size that can be processed at a single step.
This challenge can be addressed by quantum computers potentially providing better global optimality by solving larger subproblems.

\section{Data \& Methods}
\label{sec:benchmarking}
\subsection{Data}

For the purposes of benchmarking, we generate test data inspired by internal, industrially-relevant use cases. 
As described in Section~\ref{sec:applications}, the problems are represented as directed acyclic graphs, with nodes representing jobs and edges their dependencies. 
The graphs vary in size, from 5 to 30 nodes in increments of 5, and the resources associated with each job are drawn uniformly between 1 and 10.

\begin{figure}
     \noindent\centering{
    \includegraphics[width=85mm]{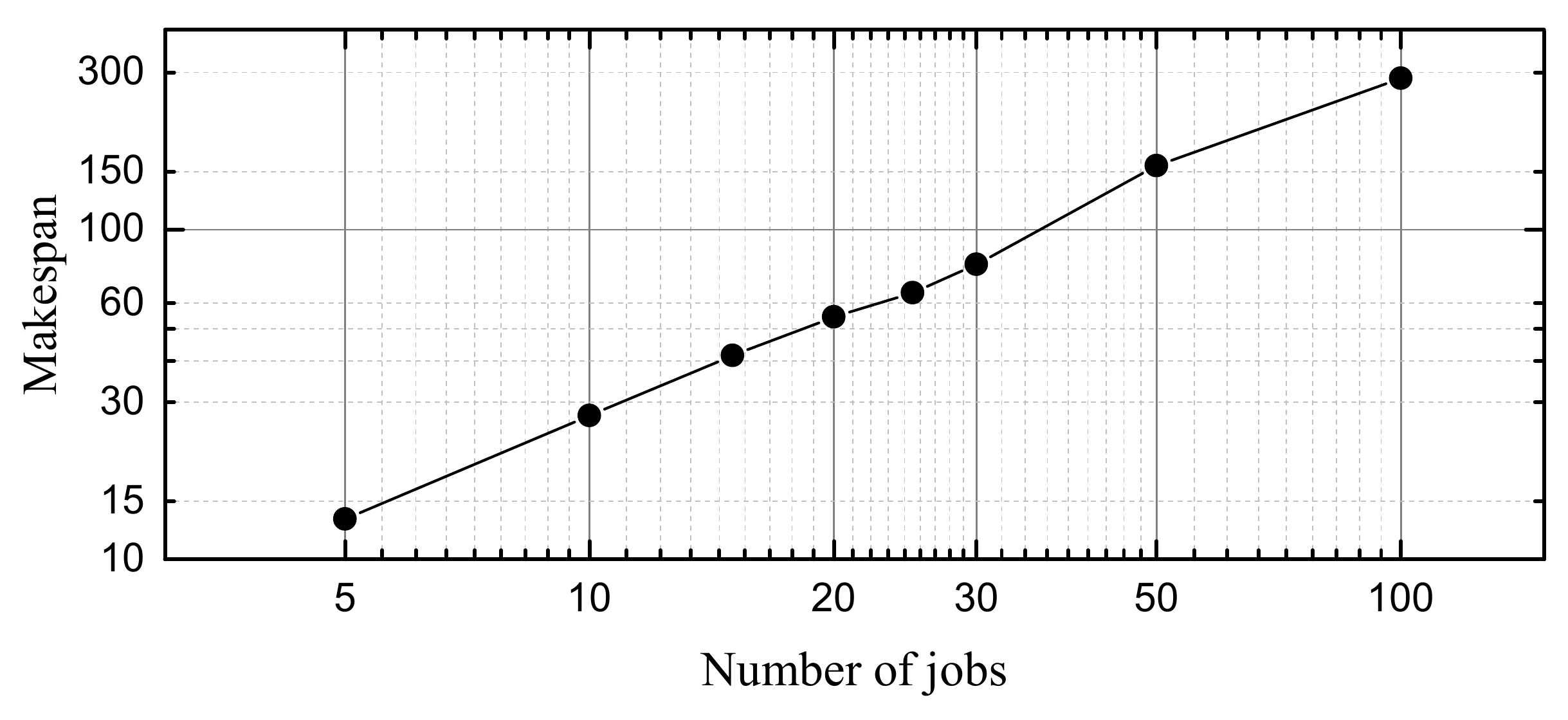}
    }
    \caption{ 
    {\bf Average makespan obtained via greedy algorithm as function of number of jobs.}
    In order to solve the problem as QUBO we need to know what is the maximum number of resource may be required -- it poses the upper bound on the expected makespan.
    This value is obtained by solving the problem with fast greedy algorithm ensuring that the problem can be solved at least with the greedy makespan.
    For our data, the makespan grows linearly with the number of jobs as $(2.75\pm0.06)\,N$.
    }
    \label{makespan}
\end{figure}

In graphs derived from the use cases serving as inspiration for the benchmarking data set, patterns of connectivity can vary widely. 
To explore the potential effects of this variation, we generated graphs with edge probabilities drawn from different distributions. 
More specifically, graphs are generated by sequentially adding nodes until the desired problem size is reached; the probability of a node having a previous node as its parent is a function of the previous node's order in that sequence (for instance, if this function is $1/x$, the tenth node added to the graph has a $1/10$ probability of having the first node as its parent). We tested three different such parent probability distributions (or \textit{fall-off} distributions), generating sets of instances using $1/x$, $1/x^2$, and $1/\sqrt{x}$. We found that the expected densities of the problem graph, the respective QUBO, and the expected makespan did not differ significantly in the problem sizes we studied. Therefore we choose to present $1/x$ in this paper (which generated sufficiently complex graphs) and leave varied graph connectivity as a topic for future work. Results from the other fall-off probabilities were qualitatively similar.

\subsection{Algorithms}

In this section we describe various algorithms used to solve the workflow scheduling problem. We consider classical, quantum and hybrid algorithms.  \\

{\it Greedy algorithm}

The greedy approach can handle any problem size. However, by definition, it is often not optimal.
The algorithm keeps in memory all parent-children relations, and firstly schedules the roots.
After completing them, it removes these jobs from the initial graph and evaluates new roots.
These new roots are then processed, removed then from the graph, and so on. This procedure is repeated until all jobs are completed.
The non-optimality of such an approach can be seen for the simple 6-node graph presented in Fig.\,\ref{5nodes}c.\\
\begin{figure*}
     \noindent\centering{
    \includegraphics[width=180mm]{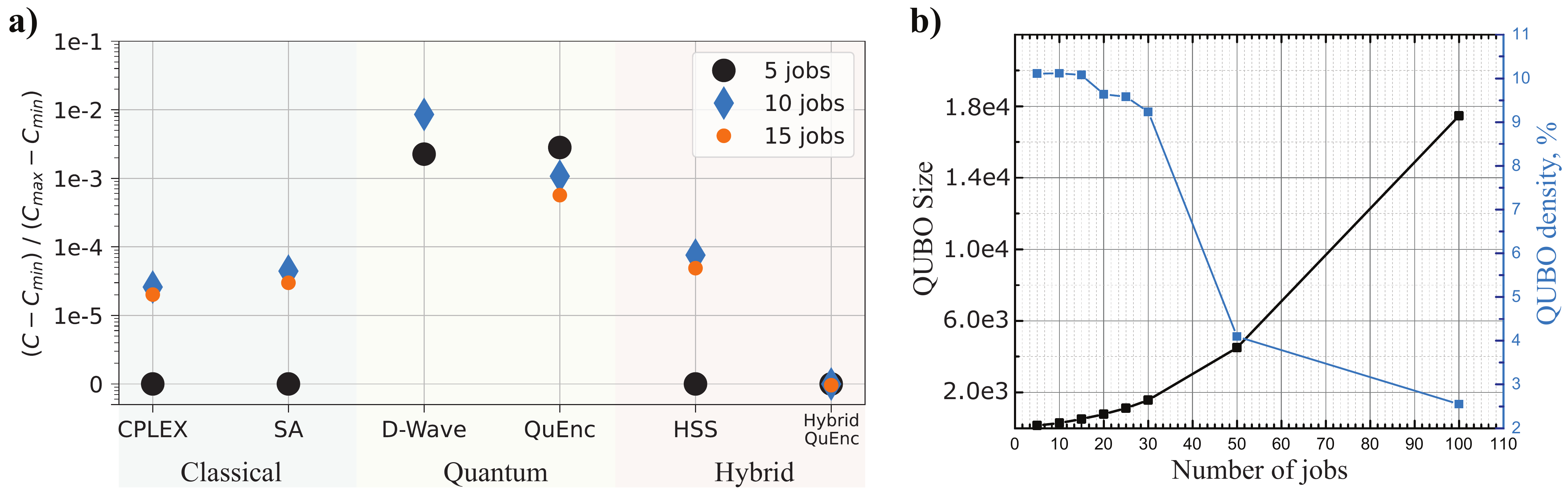}
    }
    \caption{ 
    {\bf QUBO solutions using classical, quantum, and hybrid algorithms, and problem size analysis.}
    a) The normalized cost function value for 5, 10, and 15-job scheduling problem solved as a QUBO on classical solvers, CPLEX and SA, quantum, annealing and QuEnc, and hybrid, HSS and Hybrid QuEnc based on greedy decomposition. 
    It is clear that the greedy decomposition with the QuEnc engine (that can be potentially replaced with any other suitable solver) is the only approach that can schedule jobs optimally.
    Since the maximum size of subproblems is fixed, Hybrid QuEnc can solve arbitrary large problem.
    b) Size of a QUBO as function of number of jobs $N$.
    In our data the number of required resources grows as $(2.75\pm0,06)\,N$, and the QUBO size grows $\sim N^{1.8}$.
    To schedule 5 jobs the QUBO is formulated for 60 binary variables, for 10-job problem we work with 210 bits, and for 15 we need 720 bits.
    }
    \label{results}
\end{figure*}

{\it Classical Exact Solver: Linear Programming}

Linear programming is one of the most powerful computing paradigms for discrete optimisation. 
Here, we implement the constraints described in Section\,\ref{bin_opt} in linear form, and optimize the cost function from Eq.\,\eqref{obj_cost}. We use the branching-based CPLEX solver \cite{cplex2009v12}.
This algorithm successfully finds the optimal scheduling for 30 jobs and 80 available time slots, but struggles to solve larger problems.
The runtime scaling for this problem is roughly exponential, which is not surprising since the problem class is NP-hard in the worst case, and CPLEX is an exact solver.\\

{\it Classical Exact Solver: QUBO}

To solve the QUBOs classically, we run the CPLEX as a quadratic programming solver \cite{cplex2009v12}.
Here, we limit the runtime to 10 minutes. 
The results in this case are worse than those for the LP using CPLEX, since we generate more variables in the QUBO model.\\

{\it Classical Metaheuristic Solver: Simulated Annealing}

While branching-based CPLEX is an exact solver, and so it can find the optimal solution in infinite time, metaheuristic approaches are usually used either to find sub-optimal solution quickly, or optimal solutions with some probability.
The sub-optimal solution can then be used as a starting point for an exact solver.
Here, in the framework of the QUBO, we use simulated thermal annealing (SA) as a metaheuristic QUBO solver.
We use the implementation from Ref.\,\cite{SA_Isakov2015}, a fast and robust solver written in C$++$ with a linear temperature schedule.
We set the number of sweeps to 50,000 and number of attempts to 10,000 -- a parameter setting which corresponds to 10 minutes of wall-clock time.\\

{\it Quantum Annealing Solver: D-Wave System}

In contrast to thermal annealing, quantum annealing has the potential of avoiding local minima and therefore a providing better solutions to QUBO problems.
Here we use D-Wave's Advantage quantum processing unit (QPU), which has over 5,000 qubits and 15-way qubit connectivity~\cite{Dwave}.
The limited connectivity forces us to use minor-embedding techniques to map our problem to the QPU's topology by \textit{chaining} multiple physical qubits to represent a single logical qubit. Thus, arbitrary topologies can be realized in QPUs, but with polynomial overhead in the number of qubits used to represent the problem.
For instance, one 10-job scheduling problem as a QUBO requires 210 logical qubits, but with embedding leads to 2,206 qubits, which exceeds the original size by almost an order of magnitude.
For the 15-node problem, it was impossible to find a valid embedding on the Advantage system.\\

{\it Quantum Gate-based Solver: Terra Quantum's QuEnc}

Inspired by variational quantum optimization algorithms and quantum machine learning techniques, we use the recently-proposed QuEnc algorithm~\cite{WhitePaperTQ, QuEnc}.
QuEnc is a heuristic QUBO solver for gate-based quantum systems.
Using an amplitude encoding mechanism, it is possible to encode a $n_c$-variable problem using $O(\log{n_c})$ qubits, which differs it from QAOA~\cite{farhi_quantum_2014}.
The algorithm also utilizes different ans\"atze, optimisation techniques, and circuit expressability analysis.
\\

{\it Hybrid Solver with QUBO decomposition: D-Wave HSS}

Mitigating the restrictions posed by the existing hardware, hybrid methods were proposed to decompose large problem into smaller instances so they can be solved using quantum algorithms.
One of such hybrid algorithm is the Hybrid Solver Service from D-Wave Systems, where the system finds cores of a problem, splits it into smaller pieces via classical algorithms and sends them to a quantum annealer.
Such an algorithm is not guaranteed to be optimal, but can be used as a competition metaheuristic, similar to other annealing-based algorithms.
\\

{\it Hybrid Solver with Greedy decomposition: Hybrid QuEnc} 

Here, we combine the greedy decomposition technique introduced in Section\,\ref{decomposition_section}, with the quantum engine QuEnc.
We divide the scheduling problem into subproblems, each of which is then solved via QuEnc.
It is worth noting that one can utilize any such solver to solve the subproblems, but we pick QuEnc for the potential scaling of gate-based quantum algorithms.

\section{Performance comparison}
\label{sec:results}
\begin{figure}
     \noindent\centering{
    \includegraphics[width=77mm]{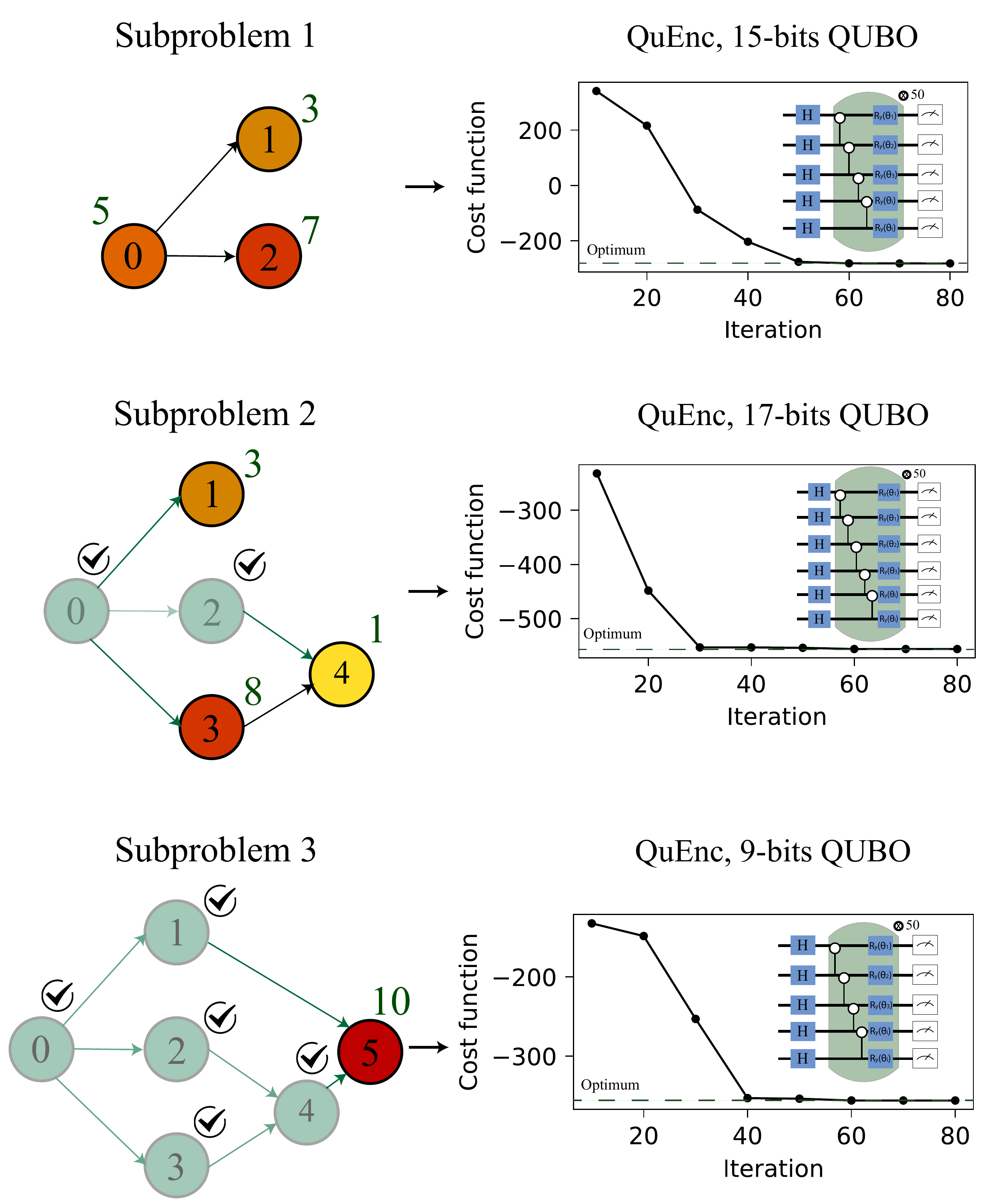}
    }
    \caption{ 
    {\bf Hybrid Quantum solution based on greedy decomposition and QuEnc algorithm.}
    The 6-node problem, decomposition of which is shown in Fig.\,\ref{decomposition}, solved via QuEnc with fixed circuit layout containing 50 layers with continuously tuned rotations.
    On the right-hand-side we show the QuEnc's convergence and corresponding quantum circuits with 5, 6, and 5 qubits that solve 15-bit, 17-bit and 9-bit QUBOs for subproblems, respectively. 
    By adjusting the QuEnc hyperparameters one can achieve faster convergence with less gates but it requires additional time-consuming tuning, which we avoid here.
    The final scheduling coincides with the optimal one provided by linear programming.
    }
    \label{hybrid_quenc}
\end{figure}

As an illustrative example, we solve three sizes of scheduling problems with 5, 10 and 15 jobs via all algorithms described above.
As a benchmark, we use the LP solution and find the minimum and the maximum cost function value of the corresponding QUBO, $C_{min}$ and $C_{max}$, that we used to normalize the cost function value. The results of the solutions comparison are depicted in Fig.\,\ref{results}a.

Among classical approaches we compare exact -- CPLEX -- and metaheuristic -- SA -- QUBO solvers.
While both CPLEX and SA optimally solve the 5-jobs problem, 10-jobs and 15-jobs scheduling can not be solved via SA.
CPLEX can find the optimal scheduling in general, but within the limited time window of 60 minutes it fails and provides only suboptimal QUBO solution.
Moreover, in the case when runtime is limited to a few seconds simulated annealing provides better solutions.

For the quantum solvers, quantum annealing is controlled by annealing time and number of samples, while QuEnc is controlled by circuit depth, learning hyperparameters and number of repetitions.
We vary the solver parameters taking into account that increasing number of reads and annealing time both leads to the higher probability to find optimal solution, but problem run duration range is limited. For The Advantage system, we used an annealing time of 2\,ms and collected 500 samples.

For QuEnc, we fix the number of layers to be 100, perform learning using gradient descent with a fixed velocity, and repeat the algorithm 10 times.
While the 5-job problem is solved with the same optimality, QuEnc provides better solution for 10-job problem and manages to solve 15-job problem exploiting just 10 qubits.
We want to emphasize that quantum annealing is performed on a real QPU, while QuEnc was simulated, nevertheless, it is clear that QuEnc utilizes much fewer resources. 

Combining classical and quantum algorithms together, we tested two hybrid solutions.
The first, HSS, successfully solves the 5-job problem, but can not schedule 10- and 15-job problems, providing less optimal solutions than simulated annealing.
As a second approach, we combined greedy decomposition with QuEnc as the subproblem solver, using 2 time slots and 3 jobs in a single subproblem.
We observed that QuEnc with 50 layers requires 5.3 restarts on average to converge to the optimal solution, with a maximum QUBO size of 17 variables for subproblems.
The example of the 5-job problem solution is presented in Fig.\,\ref{hybrid_quenc}.
On the right-hand-side we plot the simulated QuEnc convergence and corresponding quantum circuits. 
As can be seen, the circuit used in the experiment is hardware-efficient -- it utilizes connections only between neighbour qubits.
With given decomposition, 6 qubits at most are required.
Hardware-efficient circuit and small number of qubits demonstrate the possibility to run these algorithm on NISQ devices.
Further increasing the subproblem size leads to an increase in the number of qubits improving the global solution optimality for some problems, as discussed further in Section\,\ref{dec_analysis}.\\

The QUBO size as function of number of jobs is shown in Fig.\,\ref{results}b.
From the greedy solution we observed that the number of required time slots to schedule $N$ jobs scales as $M=2.8\,N$.
The QUBO size before reduction is $NM+N\log_2{r_{max}}$, where $r_{max}$ is the maximum number of resources available in a single time slot.
The first term corresponds to the decision variables (jobs and time slots), and the second to slack (ancillary) variables.
Since $M=O(N)$ and $r_{max}$ are fixed in our problem type, the QUBO size scales as $O(N^2)$.
For considered data, the QUBO size scales as $O(N^{1.8})$.
In order to estimate the weight $A$ from Eq.\,\eqref{qubo_two_terms} let us consider the case $R=0$ and $f(t)=t$ for objective function in Eq.\eqref{obj_cost}:

\begin{equation}
    \widetilde{C} = \sum_{i,t>0}tx_{i,t}\leqslant t_{\text{max}}\sum_{i,t}x_{i,t}\leqslant 2.8\,N^2.
\end{equation}

\subsection{Decomposition analysis}
\label{dec_analysis}

\begin{figure}
     \noindent\centering{
    \includegraphics[width=80mm]{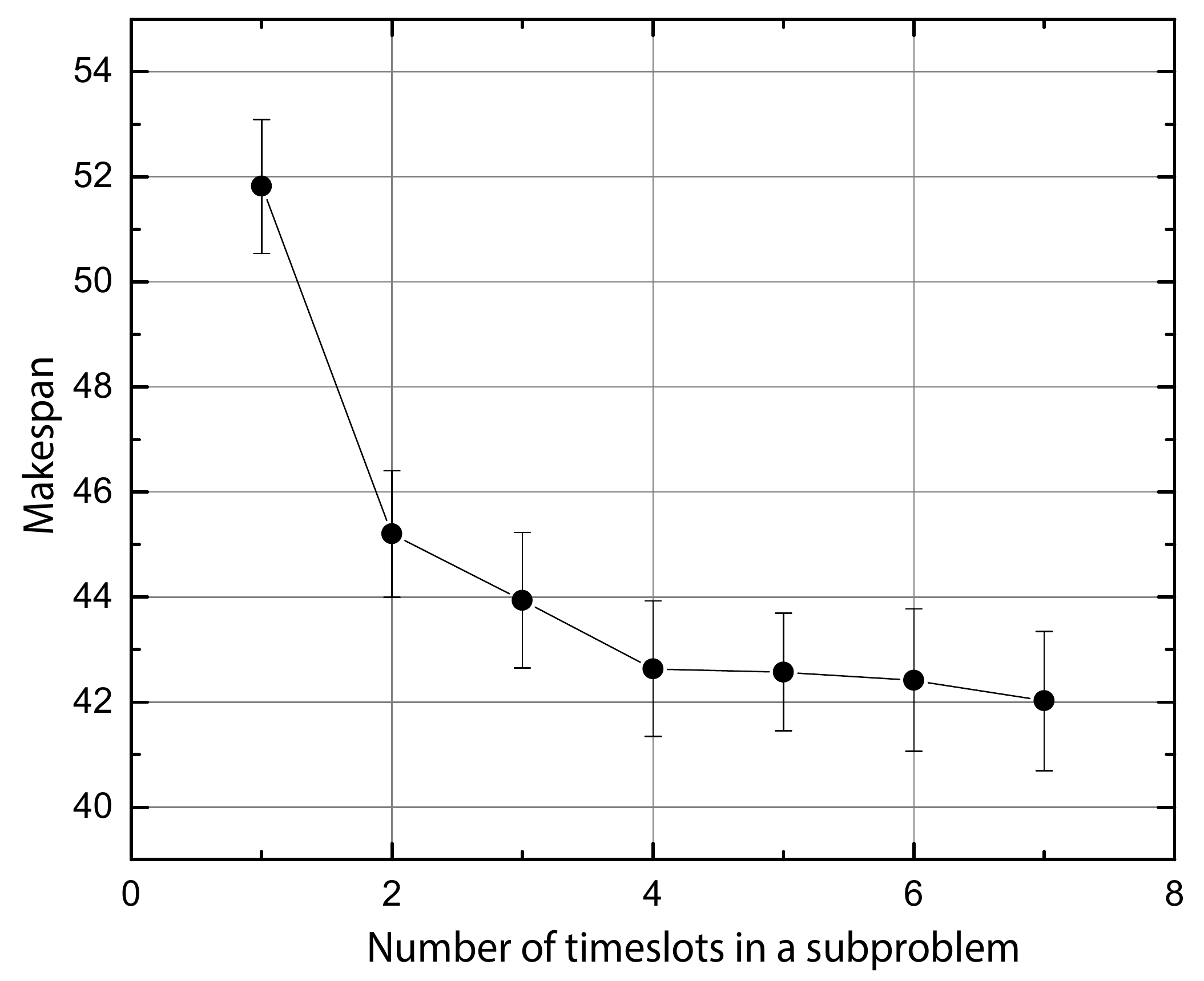}
    }
    \caption{ 
    {\bf Average makespan dependency on the size of a subproblem.}
    Applying decomposition to 50 problems from test data with 20 nodes and $1/x$ fall-off probability, we can see the predictable decreasing in average makespan with the growth of the sub-problem size. Here, number of time slots related to every step of the algorithm is equal to the maximum number of jobs taken for accommodation.
    }
    \label{decomposition_makespan}
\end{figure}

We investigate decomposing approach, proposed in Section\, \ref{decomposition_section}, using exact solution, in order to estimate the features of the pipeline in the situation when all the sub-problems are solved properly.
The reduction of makespan for 50 instances of a 20 nodes graph with the increase of a subproblem size is depicted in Fig.\,\ref{decomposition_makespan}. 
The factors, which influence the greedy decomposition, are (i) the globality of sub-problem, (ii) the number of considered jobs per time slot $r=n/m$, and (iii) disability to guarantee even the feasibility of the global solution.
Indeed, if coefficient $r$ is fixed, increasing the size of the sub-problem helps algorithm to allocate jobs more globally, thus the optimality of the final solution increases.
This phenomena is apparent in Fig.\,\ref{decomposition_makespan} (where $r=1$) and permits efficient practical usage to the algorithm.

Hyperparameters, such as the number of jobs $n$ and the number of timeslots $m$, including ratio $r$, should be tuned accordingly to considered sample alongside with the penalty weights in a QUBO formulation.

\section{Conclusion}
\label{sec:conclusions}
In this paper we presented a novel formulation of a particular class of scheduling problem, the workflow scheduling problem, as a QUBO. Inspired by a real-world use-case, we expanded upon previous known implementations of similar scheduling problems to include more realistic constraints in our problem. Specifically, we consider the case where some jobs are dependent on each other, as well as a maximum capacity of resources (at every time) which must be respected. We found that the introduction of these constraints increased the sizes of the QUBOs considerably, and so we investigated decomposition techniques in order to solve the QUBOs with the various quantum and hybrid solvers. We found that the hybrid and classical algorithms were the most successful in solving the instances, although no solver was able to solve all QUBOs at all sizes. The quantum solvers struggled to solve even the smallest problems. The improvement in performance due to the decomposition technique further highlights how the ``quantum'' difficulty in solving scheduling problems (and optimization problems in general) is more complex than the difficulty of the class of problems being solved. By reducing the problem size (and therefore the QUBO complexity), some of these limitations were able to be overcome. Therefore, future work will be dedicated to finding particular sub-classes of scheduling problems that can be more efficiently represented in QUBO form. Furthermore, novel implementation of hybrid quantum and quantum-inspired algorithms will also be investigated, to better address the QUBOs arising from such real-world instances.\\

\section*{Authors contributions}
A.A. and S.Ya. conceived of the project idea, developed the workflow scheduling problem in its presented form based on the industrial use case, and guided the work presented here. 
A.A. wrote the methods used to generate synthetic model data. S.Ya. generated the problem instances. 
A.I.P., S.Yu., and M.R.P. developed the QUBO formulation and benchmarked performance, A.I.P. and S.Yu. developed the corresponding software modules and evaluated complexity.
A.I.P. and M.R.P. developed and tested the decomposition method.
All authors contributed to the text of the paper. 

\bibliography{bib}
\end{document}